\documentclass{aa}  
\usepackage{amssymb} 
\usepackage{multirow}
\usepackage{graphicx}
\usepackage{amsmath}
\usepackage[normalem]{ulem}
\usepackage{booktabs}
\usepackage{tablefootnote} 
\usepackage{txfonts}
\useunder{\uline}{\ul}{}
\usepackage{ulem} 
\usepackage{float}
\usepackage{hyperref}
\usepackage{bm}
\newcommand*{\B}[1]{\ifmmode\bm{#1}\else\textbf{#1}\fi}
\makeatletter 
\def\@makefnmark{\hbox{\@textsuperscript{\normalfont\@thefnmark)}}}
\makeatother
\newcommand{\BP}{$\beta$~Pic\,}
\newcommand{\KIC}{KIC~3542116\,}

\begin{document} 

\titlerunning{Numerical Simulations of Exocomet Transits}
\authorrunning{Luk'yanyk et al}

   \title{Numerical simulations of exocomet transits: Insights from $\beta$~Pic and KIC~3542116}

    \author{I. Luk'yanyk\inst{1}\fnmsep\thanks{Corresponding author, \email{iluk@knu.ua}}
          \and
            I. Kulyk\inst{2}
          \and
            O. Shubina\inst{2,3}
          \and
          Ya. Pavlenko\inst{2,4}
          \and
            M. Vasylenko\inst{2}
          \and
            D. Dobrycheva\inst{2}
          \and
            P. Korsun$^{2,\dag}$
          }

   \institute{Astronomical Observatory of Taras Shevchenko National University of Kyiv, 3 Observatorna St., Kyiv, 04053, Ukraine
         \and
             Main Astronomical Observatory of National Academy of Sciences of Ukraine, 27 Akademika Zabolotnoho St., Kyiv, 03143, Ukraine
         \and
             Astronomical Institute of Slovak Academy of Sciences, 059 60 Tatransk\'{a} Lomnica, Slovak Republic
        \and
             Instituto de Astrofısica de Canarias (IAC), Calle Vıa Láctea s/n, E-38205 La Laguna, Tenerife, Spain
             }

   \date{Received 2023; accepted 2023}

 
  \abstract
   {In recent years, the topic of existence and exploration of exocomets has been gaining increasing attention. The asymmetrical decrease in the star's brightness due to the passage of a comet-like object in front of the star was successfully predicted. It was  subsequently confirmed on the basis of the light curves of stars observed by \textit{Kepler} and TESS orbital telescopes. Since then, there have been successful attempts to fit the asymmetrical dips observed in the stars' light curves utilizing a simple 1D model of an exponentially decaying optically thin dust tail. In this work, we propose fitting the photometric profiles of some known exocomet transits based on a Monte Carlo approach to build up the distribution of dust particles in a cometary tail. As the exocomet prototypes, we used the physical properties of certain Solar System comets belonging to the different dynamical groups and moving at heliocentric distances of 0.6\,au, 1.0\,au, 5.0\,au, and 5.5\,au. We obtained a good agreement between the observed and modeled transit light curves. We also show that the physical characteristics of dust particles, such as the particle size range, the power index of dust size distribution, the particle terminal velocity, and distance to the host star affect the shape of the transit light curve, while the dust productivity of the comet nucleus and the impact parameter influence its depth and duration. The estimated dust production rates of the transiting exocomets are at the level of the most active Solar System comets.}
   \maketitle
\section{Introduction}
Modern theories of planetary system formation predict a large population of planetesimals and dust inside circumstellar disks, which have been observed around dozens of pre-main and main sequence stars \citep{The1994, Grady1998, Habing2001}. Direct disk images have revealed complex structures, probably associated with the perturbations due to hidden planets \citep{Beust2000, Kalas2000, Lagr2020}. It has been shown that gravitational interactions between a planet and planetesimals, specifically the mean-motion resonance mechanism, can trigger a stream of star-grazing bodies \citep{Beust1996}. Such events have been observed spectroscopically as falling evaporating bodies (FEBs) \citep{ferlet1987,1990A&A...236..202B,1991A&A...241..488B,1998A&A...338.1015B, Vidal-Madjar1994, 2012PASP..124.1042M, 2013PASP..125..759W, Kiefer2014, Eiroa2016, Rebollido2020} and photometrically as dimming events with asymmetrical profiles in the photometric time series \citep{Lecavelier1995, Boyaj2016, 2017A&A...608A.132K, Kennedy2019, zieba2019, Pavlenko2022, Lec2022, Kiefer2023b, Kiefer2023a}. The asymmetric decrease in the star’s brightness when a comet-like body moves in front of the star was successfully predicted by \citet{desEtangs1999} and the library of the modeled comet transits depending on orbital parameters of the transiting bodies was calculated \citep{desEtangs1999b}. Meanwhile, sporadic spectroscopic features caused by FEBs in the spectra of \BP were fitted to retrieve some orbital characteristics of the transiting bodies and support the exocomet hypothesis for the observed phenomena \citep{Kennedy2018, wyat18}. Another interpretation of the asymmetric dimming event in the KIC\,125575548 light curve was proposed by \citet{Rappaport2012} and \citet{brogi2012} as a result of the disintegration of planet KIC\,1255b, producing an elongated dust cloud. 

The indirect support for possible similarity in the composition of Solar System comets and exocomets results from recent observations of the cometary gas in the circumstellar disks around main sequence stars, "polluted" white dwarf atmospheres, numerous spectroscopic observations of the FEBs events (see overview published by \citet{2020PASP..132j1001S}). Moreover, the near-infrared observations of protoplanetary disks and some star formation regions show the presence of the solid stage of high volatile species located in the ice mantle, which is in agreement with the composition of Solar System comets \citep{Dartois2002, Thi2002, Cochran2015, Rocha2024}. The traces of cometary dust found in the infrared spectra of young stars \BP, HD\,31648, HD\,163296 resemble the spectral emission features of comets C/1995\, O1\,(Hale-Bopp) and C/1991\, L3\,(Levy) very closely \citep{Sitko1998}. 

For the first time, \citet{desEtangs1999} performed the numerical simulation of a stellar occultation by an exocomet based on the physical parameters of cometary dust, which were known from the investigations of Solar System comets. However, the asymmetrical dips in the star light curves observed with the \textit{Kepler} and the Transiting Exoplanet Survey Satellite (TESS) have been simulated with simple 1D models of the exponentially decaying distribution of optically thin dust in the comet tail (see, e.g., \citet{brogi2012, Rap2018, zieba2019}). In this study, following \citet{desEtangs1999}, we fit the profiles of some already known exocomet transits using the model distribution of dust particles in the coma of a comet transiting the star's disk. The set of input parameters for the coma model is based on the properties of some known Solar System comets. This approach allows some constraints on the physical properties of an exocomet dusty atmosphere: the size of particles that likely make up the exocomet comae, the power index of differential particle size distribution, the time span for the comet tail formation, and the dust production of the comet nucleus. The paper is organized as follows: the description of the data used and the data reduction is presented in Section \ref{sec:Data}. We provide a concise overview of the Monte Carlo approach to retrieve the dust distribution in the cometary coma and transit profile modeling in Section \ref{sec:Model}. Section \ref{secRes} comprises the main results of this work and discussion. The concluding remarks are summarized in the last section.

\section{Data} \label{sec:Data}
In this study, we analyze the time series data for two stars, \BP (TIC270577175, HD\,39060) and KIC\,3542116, gathered by the TESS and \textit{Kepler} orbital telescopes, respectively. Both stars were extensively investigated and the signatures of exocomet transits have been identified in their light curves (\citep{Rap2018, zieba2019, Pavlenko2022, Lec2022}). We selected these stars because their light curves exhibit multiple transit events of the different durations and depths.

\begin{figure}[t!]
        \centering
        \includegraphics[width=1.0\linewidth]{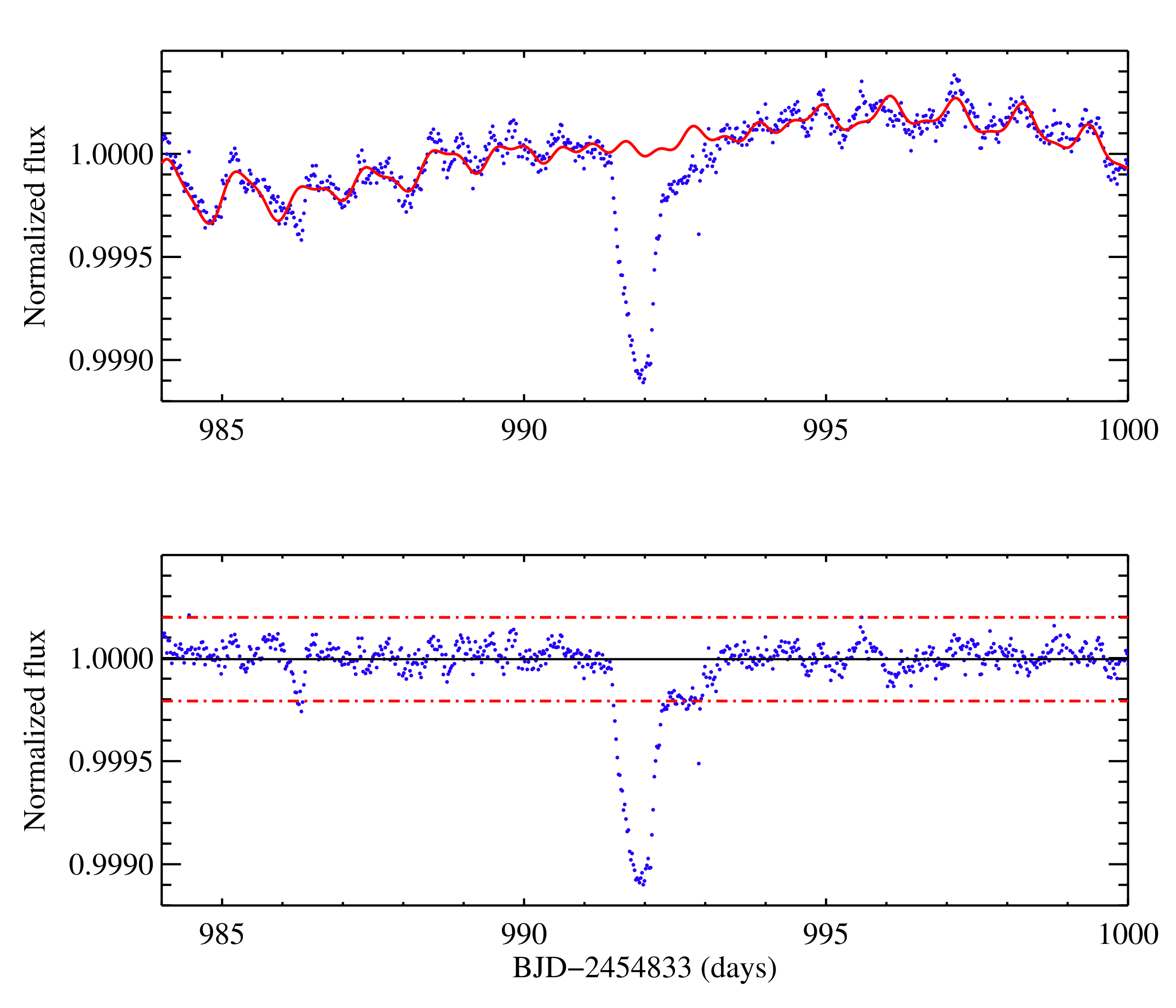} 
        \caption{Segment of the KIC3542116 light curve with the deepest exocomet transit.
{\it Top:}  Normalized PDC\_SAP flux with the harmonic content (blue dots), while the red line indicates the modeled harmonic fluctuations. {\it Bottom:} Difference between the original and modeled flux. Dash-dotted lines represent 3$\sigma$ threshold.}
        \label{Fig:quarter10}
\end{figure}

For our analysis, we employed the TESS 2-minute short-cadence Presearch Data Conditioning (PDC\_SAP) light curves for \BP and the \textit{Kepler} long-cadence 30-minute light curves for KIC\,3542116, generated by the science analysis pipeline of the Science Processing Operation Center (SPOC; \citet{jenkins2017}). The PDC segment of the SPOC applies a series of corrections to the light curves, removing instrumental artifacts, isolated outliers, and correcting the fluxes for aperture effects, such as field crowding or the fractional loss of target flux due to star centroid drifting \citep{jenkins2016}. The light curves of both stars exhibit complex harmonic content that we need to remove from the input signal (\citet{Koen2003, Zwintz2019, Rap2018, zieba2019}). Therefore, we extracted the harmonic oscillations from the light curves using the Python package SMURFS \footnote{\tiny{https://github.com/MarcoMuellner/SMURFS}}, which is designed to identify and remove significant frequencies from a time series in a fully automated manner. From each light curve, we cut out a segment around the transit between the ingress and egress points and model the PDC\_SAP flux (sectors 5 and 6 of the TESS data for \BP and quarters 1, 8, 10, and 12 of the \textit{Kepler} data for \KIC) as a superposition of the pulsation frequencies with amplitudes down to 0.02~millimagnitude and a signal-to-noise ratio (S/N) greater than or equal to 4.0. 

\begin{figure}[t!]
        \centering
        \includegraphics[width=1.0\linewidth]{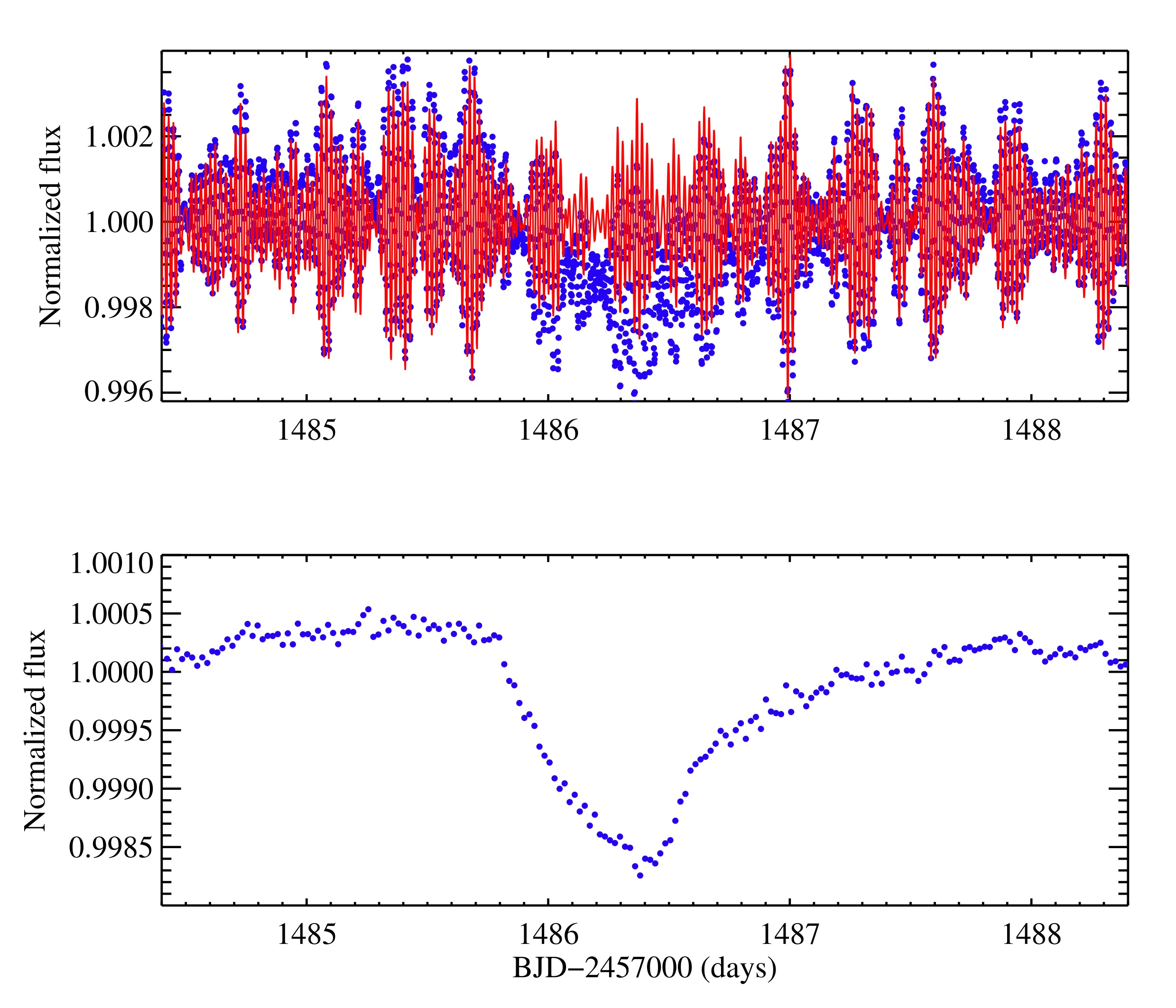}
\caption{Segment of the \BP light curve with the deepest exocomet transit.
{\it Top:} Similar to the previous figure, it shows the normalized PDC\_SAP flux with the harmonic content (blue dots), and the red line represents the modeled harmonic fluctuations.
{\it Bottom:}  Difference between the original flux and the modeled one binned in 30-minute intervals.}
\label{Fig:sector}
\end{figure}

Figures \ref{Fig:quarter10} and  \ref{Fig:sector} display the snippets of the \KIC and \BP light curves, respectively, with the deepest exocomet transits found in their light curves. The top panels of both figures show the normalized PDC\_SAP flux with the modeled flux superimposed. The bottom panels represent the difference between the observations and modeled oscillations. Dash-dotted lines shown in Figure \ref{Fig:quarter10} mark the 3$\sigma$ threshold after subtracting the star oscillations from the light curve. The 1$\sigma$ value, which equals 67.5 part per million (ppm), agrees with an estimated combined differential photometric precision (CDPP) value of 63.58\,ppm for 13-hour transit duration or 72.97\,ppm for 6 hour transit. The latter parameter is the depth estimated for a transit of a certain duration magnitude that can be registered in a whitened light curve at a 1$\sigma$ level \citep {Chris2012}. The shallow feature can be seen in the bottom panel of Figure \ref{Fig:quarter10} at approximately 986\,d, which was not mentioned in \citet{Rap2018}. However, as it surpasses the 3$\sigma$ threshold and has a duration of about 8\, hours we treated this feature as shallow transit and fitted it accordingly (details provided below). Finally, after the removal of the harmonic variations from the light curves the rms errors are between  53\,--\,60\,ppm and 170\,--\,180\,ppm for the \textit{Kepler} quarters 1, 8, 10, 12, and  TESS  sectors 5 and 6, respectively.

\section{Model} \label{sec:Model}
\subsection{Dust distribution in a cometary tail} \label{subsec:modKorun}
We modeled an exocomet transit incorporating the known physical parameters of Solar System comets. For this reason, the Monte Carlo formalism is used to simulate the dust distribution in a cometary coma, which allows for the determination of the initial conditions of dust leaving the comet nucleus before observations \citep{cashwell1959, Korsun2003, Korsun2010}. Additionally, we used a numerical code based on the Finson-Probstein theory to calculate trajectories of multiple test particles populating the dusty coma \citep{Finson1968}. The particles are assumed to be made of "dirty" ice \citep{Sekanina1975, Mukai_1986, Mukai1989, Gunnarsson_2003}. They are similar to material 1 described by \citet{Gunnarsson_2003}, consisting of a silicate core surrounded by an organic component, both covered by a mantle of "dirty" crystalline water ice with carbonaceous inclusions. We considered particles with variable masses that shrink due to the evaporation of the outer layers when a comet is approaching its parent star \citep{Korsun2010}. The differential size distribution of particles leaving the nucleus surface is assumed to follow a power law that takes the form of $dn\,\sim\,a^\textnormal{-n}da$, where $a$ is the particle radius and $n$ is the power exponent, which is considered as a model parameter \citep{Jockers1997, Fulle2004}. The proposed model of the dust tail formation does not consider nucleus rotation and circumnuclear collision zone where the gas-dynamical dragging of dust particles is significant. In addition, the model assumes that the nucleus is an isotropic source of dust and, consequently, the outflow direction of the particle has a uniform probability. According to the Monte Carlo principle, the time when an individual dust particle is released from the nucleus, $t_\textnormal{i}$, can be determined from $t_\textnormal{i}$\,=\,$R_\textnormal{i}$ $\cdot$ ($T_\textnormal{o}$\,-\,$T_\textnormal{s}$). Here, $R_\textnormal{i}$ is a random number uniformly distributed in 0 $\leq$ $R_\textnormal{i}$ < 1, $T_\textnormal{o}$ is the time of observation, and $T_\textnormal{s}$ is the time when we start to build the dust tail. Accordingly, $T$\,=\,($T_\textnormal{o}$\,-\,$T_\textnormal{s}$) is the time interval of tail formation. 

We treated the trajectory of each particle using a computer program based on the Finson-Probstein method \citep{Finson1968}. Since we are considering the non-collision zone of a comet coma, the motion of a dust particle is mainly controlled by the ratio of two forces: $\beta$\,=\,$F_\textnormal{G}$/$F_\textnormal{R}$. Here, $F_\textnormal{G}$ is the solar gravity attraction and $F_\textnormal{R}$ is the force, caused by solar radiation pressure. To calculate $\beta$ we use the simplified expression as it is adopted in \citet{Korsun2003} taking into account the physical properties of the host stars \citep{Rap2018,Arnold2019, zieba2019}:

\begin{equation} 
       \label{eq:1_00}
       \beta\,=\,0.57\frac{L_\star/M_\star}{L_\odot/M_\odot} \left(\frac{a}{1{\mu }m}\right)^{-1}
.\end{equation}

As noted above, the model considers the motion of dust particles out of the collision zone; therefore, it is more convenient to deal with terminal velocity attained by a particle at a distance of 5\,--\,20\,km from the nucleus \citep{Fink2012}. Using the gas-dynamical theory, \citet{Sekanina1992} derived the expression: $v(a)$\,=\,1/($A_\textnormal{0}$\,+\,$B_\textnormal{0}\sqrt(a))$, where $A_\textnormal{0}$ and $B_\textnormal{0}$ generally depend upon the sublimation rate,  thermal gas velocity,  nucleus mass, and the dust-to-gas-mass ratio. For the numerical calculations, we adopted the modified Sekanina’s relation \citep{Sekanina1992} derived for the dust environment of the distant comet 29P/Schwassmann-Wachmann:

\begin{equation} 
        \label{eq:1_01}
        v_\textnormal{d}(\beta)=\frac{A_1}{0.12+\beta^{-0.5}}r_\textnormal{h}^{-0.5}, 
\end{equation}
where the $\beta$ dependence of the ejection velocity, $v_\textnormal{d}(\beta),$ is used, instead of $v_\textnormal{d}(a)$. The commonly used heliocentric distance dependence of $v_\textnormal{d}$\,$\sim$\,$r_\textnormal{h}^{-0.5}$ \citep{Delsemme_1982} and an additional numerical coefficient $A_\textnormal{1}$ are incorporated, as a model parameter.
 
To reproduce the light curve of a star during an exocomet transit, it is convenient to simulate the cometary coma in the non-inertial cometocentric coordinate system. The coordinates $\xi$ and $\eta$ are in the orbital plane, directed radially outward from the star and opposite to the comet's motion along its orbit, respectively. The third coordinate $\zeta$ is in the direction perpendicular to the orbital plane, however, we are considering the distribution of dust particles in the coma and tail in the $\xi$-$\eta$ plane. The system of differential equations of dust particle motion includes the star's gravitational force adjusted for the radiation pressure via $\beta$, corrections for non-inertial effects (centrifugal and Coriolis forces), and the gravitational force of the cometary nucleus  \citep{Chorny, Korsun2010}. Neglecting the nucleus gravity we build up the distribution of dust particles in the comet coma, which allows us to calculate the extinction of the star's light at the observer direction, taking into account the transit impact parameter.

We calculate a set of 2D models of the comet dust environment based on the orbital characteristics of three Solar System comets belonging to the different dynamical groups: 1P/Halley (hereafter 1P), C/1995 O1\,(Hale-Bopp), hereafter C/1995 O1, and C/2006 S3\,(LONEOS), hereafter C/2006 S3.Table \ref{appendix:table.A0} lists the orbital parameters of the comets from the MPC data base \footnote{\tiny{https://minorplanetcenter.net/iau/MPCORB/AllCometEls.txt}}, which are used to calculate trajectories of dust particles.   
\begin{table}[h!]
\renewcommand{\tabcolsep}{0.08cm}
\begin{minipage}{\textwidth}
\caption{Orbital parameters of the selected comets}
\label{appendix:table.A0}
\begin{tabular}{lcccccc}
\hline
Comet & $e$\footnote{orbital eccentricity} & {$q$\footnote{perihelion distance}} & {$\omega$\footnote{argument of perihelion}} & {$\Omega$\footnote{longitude of the ascending node}} & {$i$\footnote{orbital inclination}} & Epoch\footnote{epoch of the last perihelion passage} \\ 
 & & (au) & ($^{\circ}$) & ($^{\circ}$) & ($^{\circ}$) &  \\
\toprule
{1P} & 0.966 & 0.60 & 112.36 & 59.43 & 162.24 & 1986/02/26.181  \\
\midrule
{C/1995 O1} & 0.994 & 0.89 & 130.17 & 282.74 & 89.25 & 1997/03/30.940 \\  \midrule
{2006/S3}\footnote[7]{dynamically new comet on hyperbolic orbit} & 1.002 & 5.16 & 140.54 & 38.62 & 166.04& 2012/04/16.386 \\  \bottomrule
\end{tabular}
\end{minipage} 
\end{table}
The start and end moments of the tail formation are the input parameters of the model, which are defined as arbitrary choices of the comet's true anomaly. Then the orbital parameters provide the comet astrocentric distance, angular rate, and angular acceleration relative to the Sun to solve the system of differential equations of the particle motion.    
The key variables for the simulation of the dust environment are the minimum and maximum size of particles populating the coma, the power exponent of the differential particle size distribution, the particle's age (the time span for the tail formation process), and the particle terminal velocities. The input parameters govern the coma's appearance, for example, the length of the dust tail and its orientation strongly depend on the tail formation time, and the particle velocities strongly influence the width of the tail in the orbital plane. The power index of the differential size distribution controls the distribution of particles of different sizes along the tail. If smaller particles are removed from the calculations, the tail tends to become shorter (for details, see \citet{Korsun2003}). To put boundaries on the model input parameters we take into account the published results of the numerical modeling of the comets considered here as well as other comets at various distances from the Sun \citep{Sekanina1992, Fulle2000, Vasundhara2002, Korsun2003, Korsun2010, Moreno2012}. The sets of the input parameters and their ranges accepted for each comet are presented in Table \ref{appendix:table.A1}. 
\begin{table}[h!]
\renewcommand{\tabcolsep}{0.04cm}
\caption{Input parameters for the Monte Carlo model used to simulate dust distribution.}
\label{appendix:table.A1}
\begin{tabular}{lccccccc}
\hline
Comet & $N_\textnormal{s}$ & $r_{\textnormal{h}}$ & $T$ & $a_{\textnormal{min}}$ & $a_{\textnormal{max}}$ & $n$ & $v_{\rm{d}}$ \\ 
  &   & (au) & (month) & (\textmu m) & (\textmu m) &   & (km$~$s$^{-1}$) \\
\toprule
{1P} & 6 & {0.6} & 1 & 0.5--0.1 & 10.0--20.0 & 2.0 -- 3.5 & 0.5--1.0 \\
\midrule
{C/1995 O1} & 5 & {1.0} & 1--2 & 0.1 & 10.0--20.0 & 2.0 -- 3.5 &0.5--3.0 \\ \midrule
{C/1995 O1} & 7 & {5.0} & 24--48 & 5.0 & 99.0--999.0& 2.0 -- 3.5&0.5--3.0  \\ \midrule
{2006/S3} & 4 & {5.5} & 24--48 & 5.0& 999.0& 2.0 -- 3.5& 0.5--3.0\\
\bottomrule
\end{tabular}   
\end{table} 

Here, $N_\textnormal{s}$ is the number of the simulation runs, $r_\textnormal{h}$ is the distance between the comet and the host star at the end moment of the tail formation, $T$ is the time span of the tail formation process, $a_\textnormal{min}$ and $a_\textnormal{max}$ are the lower and upper limits on the particle sizes, $n$ is the power index for the differential particle size distribution, and $v_{\rm{d}}$ is the particle terminal velocity. The initial total number of test particles used in the calculation is 10$^8$. The tangential component of the orbital velocity gives the speed at which a comet passes in front of the host star. It is calculated with the following equation (see Eq. 59 in \citet{Klacka1992}): 

\begin{equation} 
 \label{eq:tang_vel}
        v_\textnormal{t}\,=\,\sqrt{\frac{\mu }{r_\textnormal{h}}{(1+e\cos{\phi}})}. 
\end{equation}
Here, $\mu$, the gravitational parameter, is the product of the gravitational constant and the stellar mass;  the astrocentric  distance of the comet, $r_\textnormal{h}$, and the comet true anomaly angle at the end moment of the tail formation, $\phi$, are calculated in the process of solving the system of equations of the dust particle motion.
\begin{figure*}
        \centering
        \includegraphics[width = 0.95\linewidth]{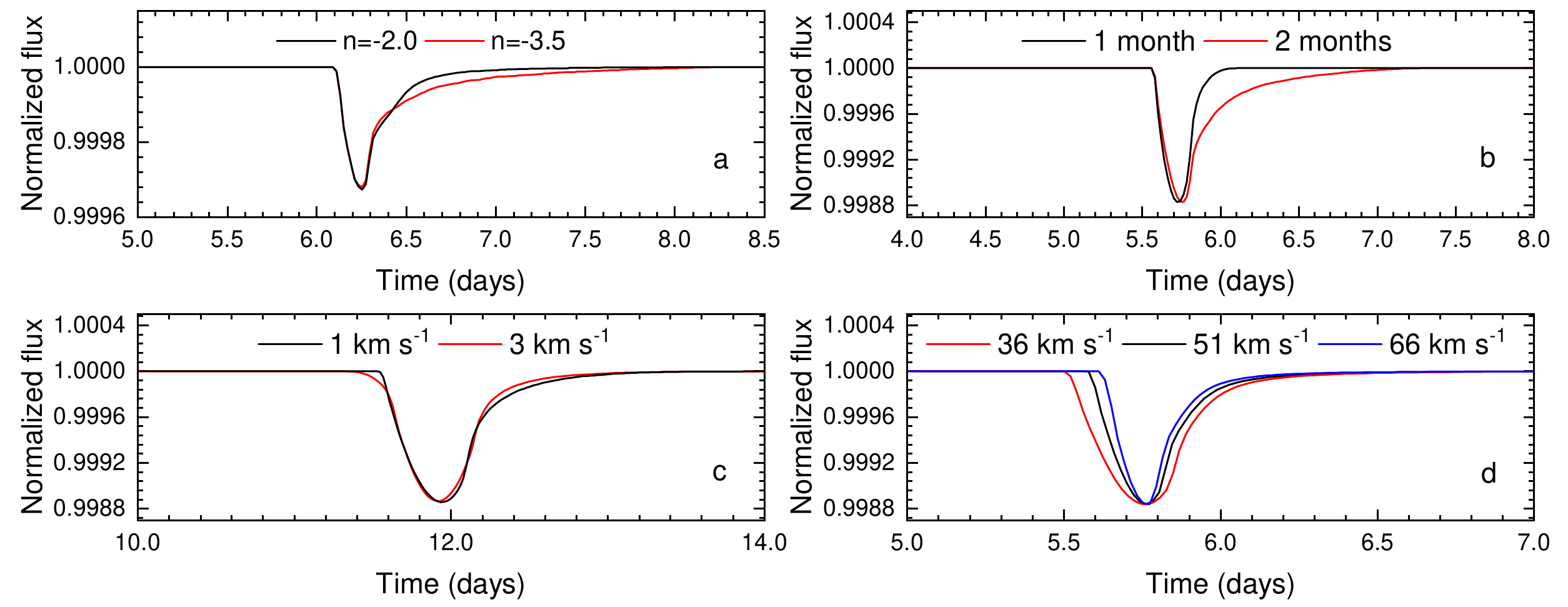}
        \caption{Dependence of the transit profile on the input parameters of the coma model listed in Table \ref{appendix:table.A1}. In all cases, the comet passes through the center of the star's disk (impact parameter $b$\,=\,0).  \textit{Panel a}: Transit of comet 1P/Halley (the index of the differential particle size distribution ($n$) changes, other parameters are fixed at: $T$\,=\,1~month; $a_\textnormal{min}$\,=\,0.1~\textmu m and $a_\textnormal{max}$\,=\,20.0~\textmu m; $v_\textnormal{d}$\,=\,1.0~km~s$^{-1}$; $v_{\rm{tang}}$\,=\,67.0~km~s$^{-1}$.) \textit{Panel b}: Transit of comet C/1995 O1 (the time of tail formation $T$ changes, other parameters are fixed at: $a_\textnormal{min}$\,=\,0.1~\textmu m and $a_\textnormal{max}$\,=\,20.0~\textmu m; $n$\,=\,3.5; $v_d$\,=\,1.0~km~s$^{-1}$; $v_{\rm{tang}}$\,=\,51.0~km~s$^{-1}$.)  \textit{Panel c}: Transit of comet C/1995 O1 (the particle terminal velocity $v_{\rm{d}}$ changes, other parameters are fixed at: $T$\,=4~years; $a_\textnormal{min}$\,=\,5.0~\textmu m and $a_\textnormal{max}$\,=\,999.0~\textmu m; $n$\,=\,3.5; $v_{\rm{tang}}$\,=\,22.7~km~s$^{-1}$.)  \textit{Panel d}: Transit of comet C/1995 O1 (the tangential velocity $v_{\rm{tang}}$ changes, other parameters are fixed at: $T$\,=2~months; $a_\textnormal{min}$\,=\,0.1~\textmu m and $a_\textnormal{max}$\,=\,20.0~\textmu m; $n$\,=\,2; $v_\textnormal{d}$\,=\,3.0~km~s$^{-1}$.)}
        \label{Fig-Prof}
\end{figure*}

Figure \ref{Fig-Prof} shows the dependence of the transit profile on the input parameters, which are used to simulate the dust distribution in the coma of the transiting comet. In all cases, the comet passes through the center of the star (impact parameter $b$\,=\,0). As an example, we select comets 1P/Halley and C/1995 O1 (Hale-Bopp) to run the models with all parameters fixed except for one, which was allowed to change within certain limits. Then, using the method described in the next subsection we estimate the change in the host star flux during the transit event. As can be seen, changes in the model input parameters as well as the comet tangential velocity affect the transit profile to varying degrees. It is expected that the comet tangential velocity influences the transit duration and, meanwhile, the coma formation time and power index of the differential particle size distribution change the transit shape at the transit egress. We note that the dust terminal velocity has little effect on the transit profile.  

\subsection{Light curve variations due to an exocomet transit} \label{subsec:exocom}
 Dips in the photometric light curve of a star result from the interaction of the star's radiation with a dusty cloud around a comet nucleus. Typically, the reduction in radiation intensity depends on the density of dust particles and their scattering and absorption properties. In our study, we assume that both the number of particles ($N$) as well as their scattering ($Q_\textnormal{sca}$) and absorption ($Q_\textnormal{abs}$) abilities, which collectively determine the efficiency factor for the total extinction ($Q_\textnormal{ext}$), exhibit weak dependencies on their chemical composition \citep{bohren1998absorption, ivanova2016comet}. On average, the extinction efficiency ($Q_\textnormal{ext}$) is approximately equal to 2 across a range of sizes larger than 1~\textmu m (for example, see figure 10 of \citet{ivanova2016comet}). We accept $Q_\textnormal{ext}$ equal to 2 to simplify the calculation of the optical depth $\tau$. The total extinction in the line of sight is therefore calculated as a sum of the extinction caused by all dust particles:

\begin{equation}
\label{eq:1_0}
        \tau=\frac{\sum_{i=0}^{N}k_{i} Q_\textnormal{ext} \pi a^2}{S}, 
\end{equation}
where  $N$ is the total number of particles in the projected area, $a$ is the average particle size, and $S$ is the projected area of the line of sight.

\begin{figure}[b!]
        \centering
        \includegraphics[width = 1.0\linewidth]{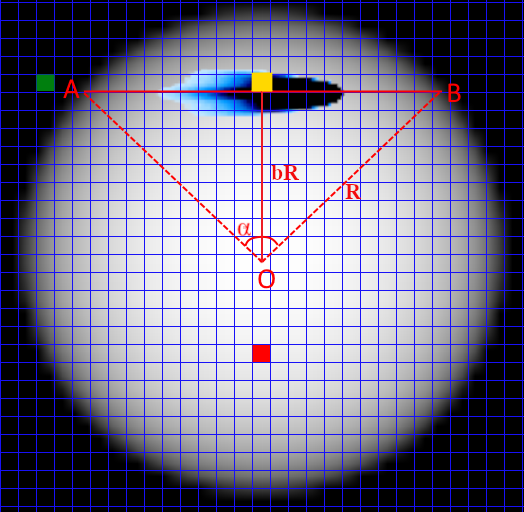}
        \caption{Visual representation of a comet transit.  $AB$ is the chord, which is the comet's trajectory across the star's disk; $\alpha$ is the central angle corresponding to the chord $AB$; $R$ is the radius of the host star; $b$\,=\,$\cos{(\alpha/2)}$ is the impact parameter.}
        \label{Fig-Geometr}
\end{figure}

To retrieve a comet transit in the star light curve, we follow the approach described by \citet{desEtangs1999}. We divide the star's surface into equal square cells, $S$, so that $\sum_\textnormal{j} S_\textnormal{j} = \pi R_\textnormal{star}^2$, with the size of each cell determined by the product of the comet tangential velocity and the mean time span between successive flux measurements. A graphical representation of a transit event is shown in Fig. \ref{Fig-Geometr}. Here, $\alpha$ is the central angle corresponding to the chord, along which the projection of the comet is moved. We also introduce the impact parameter $b$\,=\,$\cos(\alpha/2)$, which reflects how far from the center of the star the chord is located. The 2D image of the comet moving along the chord is depicted in the figure. The comet's 2D image is simulated with the method described above in Section \ref{subsec:modKorun} and represents the distribution of dust in the coma comet for a specific observing geometry. We calculated the total flux from the star each time the comet shifts to a new cell.

In this approach, the total radiation flux from the star results from the fluxes combined by two types of cells: unobscured cells (one marked with red in Fig.\ref{Fig-Geometr}) and cells eclipsed by the comet (one yellow cell in the picture as an example). The cells outside the star (green cell) are not considered. Additionally, for each cell in the star disc we consider a correction accounting for a linear limb-darkening low as $I(x,y) / I_\textnormal{c}$\,=\,1\,--\,$u$(1\,--\,$\mu$), where $I(x,y) / I_\textnormal{c}$ is the intensity of the stellar disk at the cell center (x, y) relative to the center of the star $I_\textnormal{c}$, $\mu$ represents the cosine of the angle between the line toward the cell and the normal to the stellar surface, and $u$ is the linear limb-darkening coefficient, which depends on the star parameters\citep{claret2000}. In the cited work, the authors computed the $u$ values for 12 commonly used photometric bands, 19 metallicities, and effective temperatures between 2000\,K\--\,50000\,K. Taking into account \BP and \KIC appropriate parameters, we took $u$ coefficients from the \href{http://vizier.nao.ac.jp/viz-bin/VizieR?-source=J/A\%2BA/363/1081}{"Non-linear limb-darkening law for LTE model"} catalog, which are equal to 0.275 and $u$\,=\,0.6275, respectively \citep{Rap2018, saffe2021}. 

The ratio of the star flux observed through the cometary coma to the unshadowed star flux is $F$\,=\,$F_\textnormal{ext}/F_\textnormal{star}$. To calculate this ratio, we used the approach of \citep{desEtangs1999} taking into account the limb-darkening effect:
\begin{equation}
\label{eq:1}
F = \frac{\sum_{i=1}^{N_c} S_i \cdot (1-u(1-\mu _i)) \cdot e^{-\tau_i}}{\sum_{i=1}^{N_c} S_i \cdot (1-u(1-\mu _i))}.
\end{equation}
Here, $S_\textnormal{i}$ represents the area of the $i$-th cell, $N_\textnormal{c}$ is the total number of the cells, $\tau_\textnormal{i}$ is the optical depth for a line of sight through the comet dust in the $i$-th cell, while  $u$ and $\mu$ are the limb-darkening parameters introduced above. For the unobscured cells, $\tau_\textnormal{i}$ is equal to zero and the limb-darkening effect only is taken into account. 

\begin{table*}[h!]
\begin{minipage}{\textwidth} \centering
\renewcommand{\tabcolsep}{0.14cm}
\caption{Set of the parameters providing the best fit to the observed exocomet transits.}
\label{tab:2}
\centering
\begin{tabular}{clcccccccccc}
\toprule
   \textnumero & Star/quarter & Time\footnote{moment of the flux minimum; BJD-2457000 d (\BP), BJD-2454833 d (\KIC)}) & Comet & $v_\textnormal{tang}$ \footnote{the comet tangential speed corresponding to the distance from the star noted in Table \ref{appendix:table.A1}} & $a_\textnormal{min} \footnote{the minimum and maximum sizes of particles populated the tail}$ & $a_\textnormal{max}$ & $n$  \footnote{the power index of the dust size distribution} & $v_\textnormal{d}$ \footnote{the particle terminal velocity} & $d$, \footnote{the dust coefficient} & $Q_\textnormal{d} \times$10$^{6}$ & $b$ \footnote{ the impact
parameter b} \\ 
& & (days) & & (km s$^{-1}$) & (\textmu m) & (\textmu m) & & (km s$^{-1}$) & $\times$10$^{17}$ & (kg s$^{-1}$) & \\
\midrule 
1  & \multicolumn{1}{c}{\begin{tabular}[c]{@{}c@{}} \BP \\ 5 quarter \end{tabular}}& 1442.4 & C/1995 O1 (Hale-Bopp) & 37.6$^{+0.4}_{-0.4}$ & 0.1 & 20.0 & -2.0 & 3.0 & 4.4 & 2.42  & 0.4$^{+0.2}_{-0.2}$ \\ \midrule   
2  & \multicolumn{1}{c}{\begin{tabular}[c]{@{}c@{}} \BP \\ 6 quarter \end{tabular}} & 1486.4 & 2006/S3 (LONEOS) & 16.9$^{+1.2}_{-1.2}$ & 5.0 & 999.0 & -3.5 & 0.5 & 16.4  & 0.02 & 0.6$^{+0.07}_{-0.07}$   \\ \midrule  
3  & \multicolumn{1}{c}{\begin{tabular}[c]{@{}c@{}} \KIC \\ 1 quarter \end{tabular}}& 136.1 & C/1995 O1 (Hale-Bopp)  & 51.0$^{+6.6}_{-6.6}$ & 0.1 & 20.0 & -2.0 & 3.0 & 2.2  & 1.20  & 0.3$^{+0.2}_{+0.2}$ \\ \midrule 
4  & \multicolumn{1}{c}{\begin{tabular}[c]{@{}c@{}} \KIC \\ 1 quarter \end{tabular}}& 140.0  & C/1995 O1 (Hale-Bopp) & 51.0$^{+19.9}_{-19.9}$  & 0.1  & 20.0  & -2.0    & 3.0   & 6.3 & 3.45 & 0.6$^{+0.3}_{-0.3}$ \\ \midrule 
5  & \multicolumn{1}{c}{\begin{tabular}[c]{@{}c@{}} \KIC \\ 1 quarter \end{tabular}}& 161.5  & C/1995 O1 (Hale-Bopp) & 36.0$^{+1.4}_{-1.4}$ & 0.1  & 20.0  & -2.0    & 3.0   & 2.3  & 1.25 & 0.2$^{+0.2}_{-0.2}$\\ \midrule 
6  & \multicolumn{1}{c}{\begin{tabular}[c]{@{}c@{}} \KIC \\ 8 quarter \end{tabular}}& 742.6  & 1/P Halley  & 51.1$^{+3.4}_{-3.4}$  & 0.1   & 20.0  & -2.0    & 1.0   & 3.9  & 1.71 & 0.3$^{+0.2}_{-0.2}$ \\ \midrule
7  & \multicolumn{1}{c}{\begin{tabular}[c]{@{}c@{}} \KIC \\ 8 quarter \end{tabular}}& 792.8  & 1/P Halley  & 47.0$^{+5.0}_{-5.0}$  & 0.1   & 20.0  & -2.0    & 1.0   & 7.2  & 2.66 & 0.8$^{+0.1}_{-0.1}$ \\ \midrule 
8  & \multicolumn{1}{c}{\begin{tabular}[c]{@{}c@{}} \KIC \\ 10 quarter \end{tabular}}& 986.3  & C/1995 O1 (Hale-Bopp) & 51.0$^{+8.3}_{-8.3}$  & 0.1   & 10.0  & -3.5    & 0.5   & 7.8  & 0.13  & 0.5$^{+0.2}_{-0.2}$  \\ \midrule 
9  & \multicolumn{1}{c}{\begin{tabular}[c]{@{}c@{}} \KIC \\ 10 quarter \end{tabular}}& 992.0  & 2006/S3 (LONEOS) & 16.9$^{+0.8}_{-0.8}$  & 5.0   & 999.0 & -3.5    & 0.5   & 6.8  & 0.08 & 0.4$^{+0.1}_{-0.1}$  \\ \midrule 
10 & \multicolumn{1}{c}{\begin{tabular}[c]{@{}c@{}} \KIC \\ 12 quarter \end{tabular}}& 1175.8 & 1/P Halley  & 27.0$^{+1.7}_{-1.7}$  & 0.1   & 20.0  & -2.0    & 1.0   & 9.7  & 35.30 & 0.3$^{+0.2}_{-0.2}$\\
\bottomrule
\end{tabular}
 \end{minipage} 
\end{table*}

 We sum the number of particles in each cell to calculate $\tau_\textnormal{i}$ while the comet image "moves" across the star's disk. The fixed total number of the test particles is taken as an input parameter to simulate the dust distribution over the coma. However, as follows from Equation \ref{eq:1_0}, to change $\tau_i$, we need to change $N_\textnormal{i}$ as well (if $Q_\textnormal{ext}$ is a constant); therefore, to fit a depth of a transit event (see below) we introduce the dimensionless coefficient, dust coefficient, $d$, which artificially enhances the dust productivity of the comet nucleus not distorting the particle size distribution and at the same time increasing the number of particles in each cell, $N_\textnormal{i}\times d$. The dust coefficient serves as a free parameter during the fitting of the transit profile.   

Consequently, the modeling of the transit profile involves taking the $\tau$ calculated from Equation \ref{eq:1_0} and $F(t)$ from Equation \ref{eq:1} for each time span when the simulated comet crosses the star's disk. We conducted the fitting in two steps to attain the similarity between the modeled and real transit profiles. The transit profiles were calculated for each of the simulated comet images, namely, $N$ models with the different sets of the input parameters (see Table \ref{appendix:table.A1}). At this point, we focused on selecting the models that retrieve the observed profile shape. From the modeled curves, we selected those that provide a minimum of the differences between the observed and modeled transits at their depths and half depth widths at a precision level of 1$\sigma$  (see Section \ref{sec:Data}), fixing the impact parameter, $b$\,=\,0, and varying the dust coefficient, $d$.  The dust coefficient and the speed of a transiting comet determine the transit depth and half depth width, respectively. We adopted the comet tangential speed calculated with Equation \ref{eq:tang_vel} as a first approximation. Secondly, a more precise fitting was made for the models surviving in the previous step. We re-calculated the transit profiles with the parameter $d$ fixed, varying the impact parameter $b$ to achieve a minimum of the rms error calculated from residuals between the observed and modeled fluxes within the transit interval. The impact parameter governs the chord of the comet's path along the star’s disk, hence, there is an interplay between the impact parameter and the velocity of the transiting comet. Therefore, the variation in the impact parameter results in the correction of the tangential velocity chosen at the first stage of the fitting procedure. The minimum of the rms error of residuals between the modeled and observed transit profiles determines the best fit values of the $b$ parameter and a threshold of 100\,ppm puts limits on their acceptable values and thus constrains the comet tangential velocity. The threshold is based on the fact that for the best model selected after the first step of the fitting procedure, the worst value of about 100\,ppm occurs for the $b$ parameter equal to 0.8 or more (nearly tangential transit).  

Having the best fit dust coefficient value, we calculate the particle production rate in the number of particles per unit of time for the dust ejection as $Q_\textnormal{n}$\,=\,$10^{8}$\,$\cdot$\,$d$\,$ \cdot $\,$T^{-1}$, where $T$\,--\, is the time of the tail formation. 
 
\section{Results and discussion} \label{secRes}
To illustrate the relevance of the approach proposed, we fit the transit events already found (except of the one in quarter 10 of the \textit{Kepler} data) in the TESS and \textit{Kepler} time series data. It should be noted that although all the transits have asymmetrical profiles resembling the shapes predicted theoretically by \cite{desEtangs1999b}, two morphological types mostly occur among them. The transits that produce deeper and wider dips in the star light curves, namely, deeper than 1100 ppm with a duration of more than 1.5 days between the ingress and egress points, along with other ones that cause a more shallow decrease in the stars' brightness, less than 500\,ppm with the transit duration less than 0.8 days. We fit the deep transits in sector 6 of the TESS observations of $\beta$ Pic, and in quarters 10 and 12 of the \textit{Kepler} observations of KIC\,3542116 and other shallow ones -- including a previously unidentified transit in quarter 10 of KIC\,3542116. It is important to emphasize that the fitting of the observed transit profiles involves a series of 2D images generated with various sets of physical parameters listed in Table \ref{appendix:table.A1}. However, only one of these parameter sets yields an acceptable result, producing a light curve with a shape similar to the observed transit profile at a level of rms error less than 100 ppm.

\begin{figure*}[h!]
        \centering
        \includegraphics[width = 0.9\linewidth]{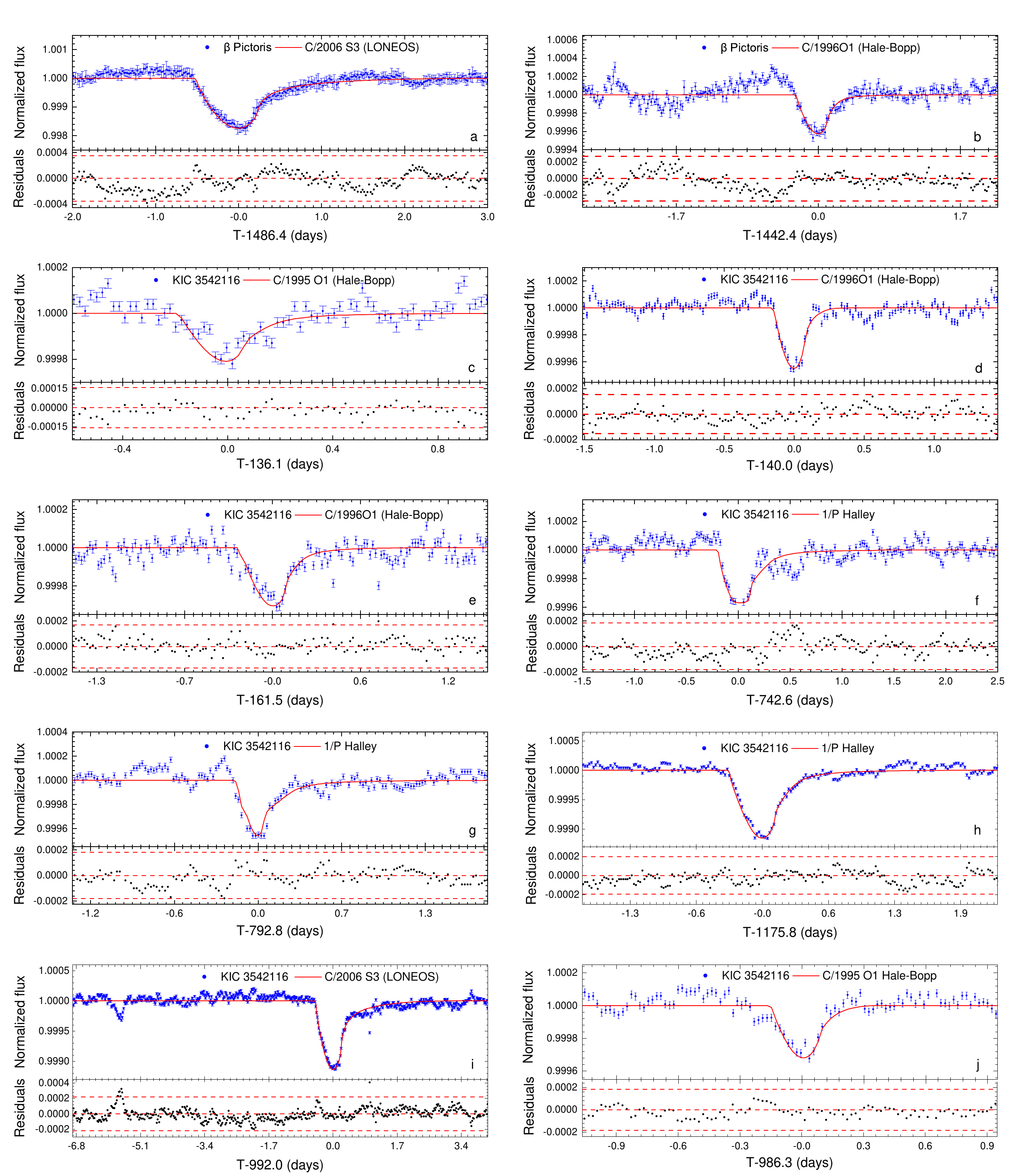}
        \caption{Segments of the \BP and \KIC light curves around the observed transit events along with the best fitted model profiles: a - \BP (sector 6) and comet C/2006 S3 (LONEOS) at r\,=\,5.5\,au; b - \BP (sector 5) and comet C/1995 O1 (Hale-Bopp)  at r\,=\,1.0\,au; c - \KIC (sector 1) and comet C/1995 O1 (Hale-Bopp) at r\,=\,1.0\,au;  d - \KIC (sector 1) and comet C/1995 O1 (Hale-Bopp) at r\,=\,1.0\,au;  e - \KIC (sector 1) and comet C/1995 O1 (Hale-Bopp) at r\,=\,1.0\,au; f - \KIC (sector 8) and comet 1P Halley at r\,=\,0.6\,au; g - \KIC (sector 8) and  comet 1P Halley at r\,=\,0.6\,au; h - \KIC (sector 12) and comet 1P Halley at r\,=\,0.6\,au; i - \KIC (sector 10) and comet C/2006 S3 (LONEOS) at r\,=\,5.5\,au; j - \KIC (sector 10) and comet C/1995 O1 (Hale-Bopp) at r\,=\,1.0\,au. \textit{Upper panels:} Blue points display the 30 min binned $\beta$ Pic normalized flux (a and b) and the normalized flux of \KIC(d-j). The red line shows the best-fit model, obtained with the 2D simulation images of respective comets (see Table \ref{tab:2}). \textit{Lower panels:} Residuals from the fit. The dashed red lines indicate the 3$\sigma$ limit. }
        \label{Fig-3_6bPic}
\end{figure*}
Table \ref{tab:2} summarizes the set of the best fit parameters for the model transit profiles presented in Figure \ref{Fig-3_6bPic}. For each simulated image of the comet, which provides the best fit, it includes the moment when the flux reaches its minimum within the transit interval, the tangential speed of the comet, the input parameters of the comet model $a_\textnormal{min}$, $a_\textnormal{max}$, and $v_\textnormal{d}$ corresponding to those in Table \ref{appendix:table.A1} as well as the dust coefficient, the dust production rate of the transiting comet (see explanation below) and the impact parameter $b$. The table also gives the upper and lower limits on the impact parameter and tangential velocity, which keep the difference between the model and observed curves within the rms error of the threshold accepted.
We transformed the particle production rate $Q_\textnormal{n}$ in particles per second into the more conventional dust mass production rate $Q_\textnormal{d}$ in kg per second (see Table \ref{tab:2}). Assuming a spherical nucleus and isotropic emission of dust, both parameters are related as \citep{Jorda1995}:

\begin{equation} 
        \label{eq:3}
        Q_\textnormal{d}=Q_\textnormal{n}\frac{4\pi}{3}\int_{a_1}^{a_2}\rho_\textnormal{dust}(a)a^3f(a)da. 
\end{equation}
Here, $a$ denotes the radius of the dust particles, $\rho_{dust}(a)$ is the size-dependent density of dust particles, and $f(a)$ is the differential size distribution. $a_\textnormal{1}$ and $a_\textnormal{2}$ are the minimum and maximum particle radii. The dust density $\rho_\textnormal{dust}(a)$ is assumed by \citet{1985AJ.....90.2591N} as

\begin{equation} 
        \label{eq:4}
        \rho_\textnormal{dust}(a) = \rho_\textnormal{0} - \rho_\textnormal{1} \Big(\frac{a}{a + \tilde a} \Big),
\end{equation}
with $\rho_\textnormal{0}$\,=\,3000\,kg~m$^{-3}$, $\rho_\textnormal{1}$\,=\,2200\,kg~m$^{-3}$ and $\tilde a$ is mean dust particle in \textmu m. This function $\rho_\textnormal{dust}(a)$ implies a higher density for small particles than for large ones and, thus, it takes the structure of interplanetary dust particles into account. We adopted the particle size distribution function $f(a)$ in the form of $a^\textnormal{-n}da$, the power index $n$ as well as  $a_\textnormal{1}$ and $a_\textnormal{2}$ are the best fit parameters to the transit profiles listed in Table \ref{appendix:table.A1}. 

In their study, \citet{Rap2018} considered the same events, under the assumption that the dips in the flux are indeed due to objects with dust tails crossing the host star. They fit these transits with the 1D model of a dust tail, which provides an exponentially decaying extinction profile. The inferred speeds of the dust-emitting body during the transits are in the range of 35\,--\,50\, km s$^{-1}$ and 76\,--\,90\, km s$^{-1}$ for the deeper and more shallow transits, respectively. These velocities correspond to the circular orbit periods of approximately 100-300\,days and 50-90\,days for the deeper and shallower transits (see Fig.11 in \citet{Rap2018}), which are much shorter than the periods of comets in the Solar System. We determined speed values of about 17\, km~s$^{-1}$  for two deepest and longest transits (sector 6 of \BP and quarter 10 of \KIC) and between  37 and 47 \, km~s$^{-1}$ for the shallower and shorter ones. Under the circular orbit assumption, these speed values correspond to orbital distances of 4.5~au and between 0.5 and 0.9~au for the deep and shallow transits, respectively. Our results are not perfectly consistent with those presented by \citep{Rap2018} likely due to the different approach to the representation of the transit events.  \citep{Rap2018} used a simple model of the comet dusty tale and the tail parameters (the optical depth just behind the comet and exponential scale length of the tail), along with the comet tangential speed and impact parameter are among the fitting ones. We simulated the comet dust environments at certain astrocentric distances taking into account the estimates on the dust physical parameters inferred from the Solar System comets. Moreover, in our case, the tangential velocities of the crossing bodies are given in advance and we use them as input values to retrieve the transit profile.

Our findings do not contradict results presented by \citet{Kiefer2014}, suggesting two distinct comet families to explain the variety of the observed FEB events. Deeper and longer transits appear to result from comets passing at greater distances from the star, such as comet C/2006\,S3\,(LONEOS) at a distance of 5.5\,au, which can explain the deep transits observed in the light curves of \BP and KIC\,3542116. Shallow and shorter transit events seem to be associated with comets similar to C/1996\,O1\,Hale-Bop at 1\,au and 1/P\,Halley at 0.6\,au. Moreover, it should be noted that the shallow and shorter transit events are approximated by the comet models with the size distribution exponent $n$\,=\,2.0, while the deep transits require a steeper distribution with $n$~=~3.5. 
The dust production rate, $Q_\textnormal{d}$, takes its values from 0.1$\times$10$^6$ to 3.5$\times$10$^6$ for shallow transits and from 0.2$\times$10$^5$ to 0.8$\times$10$^5$\,kg~s$^{-1}$ for the deep and long events, whose profiles are modeled using the simulated image of comet C/2006 S3. Highest dust production rate of 35.3$\times$10$^6$\,kg~s$^{-1}$ is estimated for the transit event in quarter 12 of \KIC data, whose depths are 1150~ppm, which is similar to the deepest ones; howeve, the duration is shorter about one~day. \citet{Rap2018} estimated the minimum value of dust mass-loss rate as $Q_\textnormal{d}$\,$\ge$\,2.5$\times$10$^{7}$~kg~s$^{-1}$ for a distance of $\sim$2.5$R_\textnormal{star}$. Except for the  event in quarter 12 of \KIC\ mentioned above, the dust production rates presented here are smaller by an order of magnitude, but they have been calculated for larger distances from the parent star. However, \citet{Kiefer2023b} estimated a dust production rate of about 0.9$\times$10$^{5}$kg~s$^{-1}$ for the exocomet transiting star HD 172555 at a distance of 1\,au. These authors also calculated the dust production rates for the FEBs observed in \BP system as between 3$\times$10$^{5}$~kg~s$^{-1}$ and 6$\times$10$^{5}$~kg~s$^{-1}$.  

Notably, most of the Solar System comets have significantly lower dust production rates. For example, C/1995 O1 (Hale-Bopp) is thought to be among the high active Solar System comets, however, its dust production rate is estimated about 5$\times$10$^3$\,kg~s$^{-1}$ at a heliocentric distance of about 5\,au,  and from 10$^5$ to 10$^2$\,kg~s$^{-1}$ in the range of heliocentric distances 3\,--\,13\,au \citep{2003A&A...403..313W}. At a heliocentric distance of 1.08\,au, comet 8P/Tuttle demonstrated activity at level of 10$^4$\,kg~s$^{-1}$\citep{10.1093/mnras/stab2609}. For comet C/2007\,D1\,(LINEAR) $Q_\textnormal{d}$\,=\, 5.30~$\times$~10$^2$\,kg~s$^{-1}$ is inferred at $r_\textnormal{h}$\,=\,9.7\,au \citep{2010A&A...513A..33M}. The dust production of active Centaur 29P/Schwassmann-Wachmann~1, which is famous for its sporadic outbursts, is estimated to be $Q_\textnormal{d}$\,=\,5.1$\times$10$^3$\,kg~s$^{-1}$ at $r_\textnormal{h}$\,=\,5.8\,au \citet{2006DPS....38.3704J}. The $Q_\textnormal{d}$\,=\,21\,kg~s$^{-1}$ is calculated for comet C/2006 S3 (LONEOS) at $r_\textnormal{h}$\,=\,5.56 au \citet{2014A&A...571A..73R}, using the same model as we did in this work and with similar input parameters but lower terminal velocity for dust particles $v_\textnormal{d}$\,=\,24.3\,m~s$^{-1}$. The dust productivity of the first interstellar comet 2/I Borisov changed from  11.2\,$\pm$\,4.4 to 16.4\,$\pm$\,7.3\,kg~s$^{-1}$ between the heliocentric distances of 1.9 and 2.5\,au \citep{2021AJ....162...97C}. It is important to highlight that comparison of the dust production rates of the different Solar System comets should be made with caution because the conversion of the observed Af$\rho$ parameters to dust production rate is accompanied by significant uncertainties, often reaching a factor of 10 \citep{2021PSJ.....2..154F}. This is attributed to the intricate light scattering process, influenced by factors such as particle size, particle size distribution, complex refractive index, somewhat uncertain particle shape, dust particle density, and scattering phase function. 

The tangential velocities of the transiting comets, as deduced from this study, reasonably agree with those calculated by \citep{Pavlenko2022,zieba2019} from the analysis of the \BP light curve, despite the different approaches used to fit the transit profiles. The tangential velocity of the comet responsible for the deepest transit in the \BP light curve is estimated to be 16.9\,km~s$^{-1}$, falling within the range of 15\,--\,17\,km~s$^{-1}$ calculated by \citep{Pavlenko2022}. For the shallow features, the transit velocities between 27 and 51\,km~s$^{-1}$ are compatible with the range of 28\,--\,55\,km~s$^{-1}$ presented by \citep{Pavlenko2022}.

The dust tail of comet C/2006\,S3\,(LONEOS) is populated by particles larger than 5\,\textmu m. Information about the properties of cometary dust beyond the water-ice sublimation zone is limited. The dust environment of distant active objects is primarily characterized using numerical models, which suggest a size range of particles between 1 and 1000\,\textmu m, low outflow velocities, and a power index of the differential size distribution between -3.2 and -4.5 \citet{Fulle1994, Korsun2003}. The simulated dust environment of C/2006\,S3\,(LONEOS), with parameters similar to distant Solar System comets, successfully reproduces the deep transits in both \BP and KIC\,3542116 light curves. The shallow transits, on the other hand, appear to result from comets that are closer to the star. Smaller grains populate their comae, likely due to a higher level of particle disintegration, suggesting the presence of small compact particles resulting in a shallower dust size distribution with a power-law index of -2 \citep{Rot2015S}.

It should be noted that the shallow transit in sector 10 (Fig. \ref{Fig-3_6bPic}, panel j) has a more symmetrical profile. The best fit for this profile is provided by the comet image simulated with the short time interval of the tail formation (see Fig. \ref{Fig-Prof}b, black line) of $n$\,=\,3.5, with the upper and lower limits on the particle sizes of 0.1 and 10 \textmu m. Such parameters imply a tail populated with short leaving fine grains. Also, we cannot rule out the idea that the peculiar profile can be attributed to specific observational geometry. As shown by \cite{desEtangs1999b},  the longitude of periastron is one of the important parameters that control the shape of a transit profile. 

\section{ Conclusions}
The observed exocomet transits in the light curves of  \BP and KIC\,3542116 could be caused by passing comet-like bodies with the dusty comae similar to those of Solar System comets C/2006\,S3\,(LONEOS), C/1995\,O1\,Hale-Bopp, and 1P/Halley.
The transits  investigated in this paper can be categorized into two distinct types, possibly caused by different comet groups:
\begin{enumerate}
 \item[i)] The shallow and shorter transits are linked to comets moving closer to the star (about 1\,au), their comae consist of dust particles sized between 0.1 and 20 \textmu m, a differential size distribution power index of -2 implies short lived fine particles populated the comet comae.
\item[ii)]  The deeper and longer dips in light curves are likely caused by comets moving at larger distances (around 5\,au). Their comae are comprised of larger dust particles, greater than 5\,\textmu m, and the differential particle size distribution is steeper (power index is -3.5).
\item[iii)] The dust production rate of the comets causing shallow transits is higher likely due to an increase in the activity level when a comet approaches the star. The estimated dust production rates exceed the values inferred for most Solar System comets and are at the level of the most active ones. It also surpasses the estimated dust production rates for some FEB events \citet{Kiefer2023b}.
\end{enumerate}

Our findings exhibit a quantitative agreement with previous results presented in \citet{zieba2019} and \citet{Pavlenko2022}, even though we employed a distinct approach to fitting the transit profiles. It is important to note that in our approach, the physical characteristics of dust particles, such as the minimum and maximum dust sizes,  power index of the particle size distribution, and the particle terminal velocity affect the shape of the transit light curve, while the dust productivity of the comet nucleus and the impact parameter influence its depth and duration.

\begin{acknowledgements}
The authors express sincere gratitude to the anonymous Reviewer for the 
thoughtful and constructive feedback, which helped us to improve the article. 
This study was performed in the frames of the government funding program 
for institutions of the National Academy of Sciences of Ukraine and 
supported by the National Research Foundation of Ukraine 
(\textnumero\,2020.02/0228). Also, the research of IL is supported by 
Project No. 0124U001304 of the Ministry of Education and Science 
of Ukraine. YP’s investigations were carried out under the MSCA4Ukraine program, 
project number 1.4-UKR-1233448-MSCA4Ukraine, which is funded by 
the European Comission.\\

All data presented in this paper were obtained from the Mikulski Archive for Space Telescopes (MAST). STScI is operated by the Association of Universities for Research in Astronomy, Inc., under NASA contract NAS5-26555. Support for MAST for non-HST data is provided by the NASA Office of Space Science via grant NNX09AF08G and by other grants and contracts. This paper includes data collected by the TESS mission. Funding for the TESS mission is provided by the NASA Science Mission directorate.
\end{acknowledgements}

   \bibliographystyle{aa} 
   \bibliography{comet} 

\end{document}